\documentclass{article}
\usepackage{ijcai25}

\usepackage{times}
\usepackage{soul}
\usepackage{url}
\usepackage[hidelinks]{hyperref}
\usepackage[utf8]{inputenc}
\usepackage[small]{caption}
\usepackage{graphicx}
\usepackage{amsmath}
\usepackage{amsthm}
\usepackage{booktabs}
\usepackage{algorithm}
\usepackage{algorithmic}
\usepackage[switch]{lineno}

\usepackage{color}
\usepackage{listings}	
\definecolor{dkgreen}{rgb}{0,0.6,0}
\definecolor{gray}{rgb}{0.5,0.5,0.5}
\definecolor{mauve}{rgb}{0.58,0,0.82}
\usepackage{tcolorbox}
\tcbuselibrary{skins}

\title{The importance of visual modelling languages in generative software engineering}

\author{
    Roberto Rossi
    \affiliations
    Business School, University of Edinburgh, UK
    \emails
    roberto.rossi@ed.ac.uk
}

\begin{document}

\maketitle

\begin{abstract}
Multimodal GPTs represent a watershed in the interplay between Software Engineering and Generative Artificial Intelligence. GPT-4 accepts image and text inputs, rather than simply natural language. We investigate relevant use cases stemming from these enhanced capabilities of GPT-4. To the best of our knowledge, no other work has investigated similar use cases involving Software Engineering tasks carried out via multimodal GPTs prompted with a mix of diagrams and natural language.
\end{abstract}

\section{Introduction}\label{sec:intro}

Software engineering (SE) applies engineering principles to the development, operation, and maintenance of software systems, ensuring they are reliable, efficient, and meet user requirements. Generative Artificial Intelligence (GenAI) models, such as pre-trained models \cite{DBLP:conf/naacl/DevlinCLT19,NIPS2017_3f5ee243} and large language models (LLMs), are revolutionising fields like computer vision and natural language processing through their ability to generate novel and contextually-appropriate content. 
In recent times, the interplay between SE and GenAI has been receiving increasing attention. A vast array of applications of pre-trained models and LLMs are surveyed in \cite{huang2024generativesoftwareengineering,10.1145/3695988}. However, when it comes to the type of data used in existing studies, it appears that all studies surveyed focused on text-based datasets, with the most prevalent type of data utilised in training LLMs for SE tasks bring programming tasks/problems expressed in natural language \cite{10.1145/3695988}. Similar conclusions are reached in \cite{huang2024generativesoftwareengineering}, which focused on seven sub-tasks of SE: requirements generation, code generation, code summarisation, test generation, patch generation, code optimisation, and code translation. In all cases, the pipeline typically begins with instructions in natural language that need to be used in the context of one or more of these seven sub-tasks. 

In SE it is often the case that communication among developers, and between developers and customers, occurs in the form of sketches and diagrams \cite{Baltes2014} and not just via natural language. The reason for this can be easily understood once we consider the following excerpt taken from \cite{tufte2003cognitive}.
\begin{quote}
A TALK, which proceeds at a pace of 100 to 160 spoken words per minute, is not an especially high-resolution method of data transmission. Rates of transmitting visual evidence can be far higher. The artist Ad Reinhardt said, "As for a picture, if it isn't worth a thousand words, the hell with it." People can quickly look over tables with hundreds of numbers in the financial or sports pages in newspapers. People read 300 to 1,000 printed words a minute, and find their way around a printed map or a 35mm slide displaying 5 to 40 MB in the visual field. Often the visual channel is an intensely high-resolution channel.
\end{quote}
It is then not surprising that the ambition of automatically turning sketches and diagrams into working code, or to reverse engineer working code into diagrams, has existed for a very long time in SE. Not only sketches and diagrams represent a high-resolution communication channel, but when they are drawn following a standard, they become a form of technical languages. The importance of technical languages is that they are denotative: they say one thing and one thing only. Conversely, natural language is connotative: the meaning of a statement is context dependent. It is the intrinsic ambiguity of natural language that makes it not well suited for programming, or for communicating programming-related matters. 

In this work, we argue that the advent of multimodal GPTs, such as GPT-4 \cite{openai2024gpt4technicalreport}, may represent a watershed in the interplay between SE and GenAI. GPT-4 accepts image and text inputs, and it can therefore receive prompts that contain sketches and diagrams, rather than simply natural language. We therefore investigate relevant use cases stemming from these enhanced capabilities of GPT-4. To the best of our knowledge, no other work has investigated similar use cases involving SE tasks carried out via multimodal GPTs prompted with a mix of diagrams and natural language.

The rest of this paper is organised as follows: 
Section \ref{sec:background} provides relevant background in GenAI and SE. 
Section \ref{sec:methodology} outlines the methodology adopted in our study. 
Section \ref{sec:GenSE} investigates the role of multimodal GPTs in software development by illustrating a selection of use cases.
Section \ref{sec:discussion} provides concluding remarks by reflecting on how the former use cases address several research gaps in Generative Software Engineering (GenSE), as well as some longstanding issues in SE, and open new research directions.

\section{Background}\label{sec:background}

This section provides relevant background in GenAI and SE.

\subsection{Software Engineering and UML}
The Unified Modelling Language (UML) is a general-purpose standardised visual language for designing systems. In SE, UML is utilised in three main areas: use case development, static analysis, and dynamic analysis of the software system. 
Use case development is a key step of requirement analysis. Use cases\footnote{User stories in Agile Software Development \cite{beck2001agile}} describe how a user interacts with a system or product to achieve a specific goal. Static analysis of a software system describes the structure of a system by showing its classes, their attributes, operations (or methods), and the relationships among objects. Dynamic analysis expresses and model the behaviour of the system over time. More specifically, in UML, use case diagrams support use case development. Class diagrams support static analysis; and interaction (sequence, activity, collaboration) diagrams support dynamic analysis. However, it should be noted that UML is a vast language that finds countless applications to go beyond the three areas here considered. 

Despite UML being the standard visual language for designing systems, recent studies suggest that that most software developers favour informal hand-drawn diagrams and do not use UML; those using UML, tend to use it informally and selectively \cite{Baltes2014}. The authors further advocated the development of suitable tools to make better use of such sketches. Our study argues that GenAI may represent such missing tool.

\subsection{Generative Artificial Intelligence}

GenAI refers to a class of AI models that can create new content, such as text, diagrams, or code, based on the patterns they learn from existing data \cite{author2023generative}. Large Language Models (LLMs) are a specific category of GenAI models that focus on language-related tasks, such as text generation, translation, summarisation, and question answering. LLMs are typically based on a neural network architecture called a Transformer \cite{NIPS2017_3f5ee243}, which is pre-trained --- hence the name Generative Pretrained Transformer (GPT) --- on a massive dataset of text and code, and which allows them to process and generate text by considering long-range dependencies and contextual information \cite{NEURIPS2020_1457c0d6}. The latest GPTs, such as GPT-4 \cite{openai2024gpt4technicalreport}, are multimodal: they accept image and text inputs, and produce image and text outputs, leading to new applications.

Prompts are instructions or input given to a large language model (LLM) to generate a specific response. Prompt engineering --- the design prompting strategies to query language models --- offers a cost-effective way to adapt pre-trained models without full fine-tuning. 

{\em Single Prompt Techniques.} 
Zero-Shot Prompting involves providing tasks with natural language without additional context, relying on the model's pre-existing capabilities. Few-Shot Prompting includes providing examples within the prompt to guide the model, enhancing its performance on complex tasks by exposing it to the input and output patterns \cite{NEURIPS2020_1457c0d6}. Chain of Thought Prompting \cite{10.5555/3600270.3602070} aids in breaking down reasoning tasks into smaller steps to improve outcome accuracy, using either zero-shot or few-shot methods to encourage step-by-step thinking. 

{\em  Multiple Prompt Techniques.} 
Voting/self-consistency \cite{wang2023selfconsistency} involves generating multiple responses and selecting the most common result, which can improve accuracy, especially for complex reasoning tasks. 
Divide and Conquer methods split tasks into subtasks handled in sequence for improved manageability and precision, seen in Directional Stimulus Prompting \cite{10.5555/3666122.3668857}, Generated Knowledge \cite{Liu2021GeneratedKP}, and Prompt Chaining. 
Self-evaluation asks the model to verify output accuracy, exemplified by Reflexion \cite{10.5555/3666122.3666499} and Tree of Thoughts \cite{10.5555/3666122.3666639}, enabling iterative improvement.
%

{\em Retrieval-Augmented Generation} (RAG)  \cite{10.1145/3637528.3671470} and ReAct \cite{yao2023react} combine LLMs with external systems to improve context handling and output relevance. 

\subsection{The role of GenAI in Software Engineering}\label{sec:literature}

In recent times, SE has become one of the important application areas for GenAI. We focus on two recent surveys \cite{huang2024generativesoftwareengineering,10.1145/3695988} investigating the interplay between SE and GenAI over a large body of recent works. 

Several studies, e.g. \cite{Arora2024,White2024}, investigated the use of GenAI in the context of the requirement engineering sub-task. GenAI does play a role in this sub-task, which is mainly concerned in turning requirements expressed in natural language by the customer into suitable user stories and/or conceptual diagrams \cite{Robeer2016}, but our focus in this study will not be on this sub-task.
Conversely, SE sub-tasks of interest in the context of the present study include: software design, software development, and code summarisation.  

Application of LLMs in software design remains relatively sparse: \cite{huang2024generativesoftwareengineering} does not include any study within this sub-task; while \cite{10.1145/3695988} only report 4 works \cite{Kolthoff2023,https://doi.org/10.48550/arxiv.2304.09181,White2024,zhang2024experimentingnewprogrammingpractice}, none of which overlaps with the content of the present study; they also stress that by expanding the use of LLMs to this under-explored area it is possible to improve how software designs are conceptualised. 

Both surveys identify a plethora of works concerned with software development (including code generation, test case generation, patch generation, and code optimisation) and code summarisation.

Code generation has been object of investigation for a long time in the AI community. Early works used symbolic and neural-semiotic approaches \cite{6679385}. However, recent neurolinguistic models, such as GPT-4 \cite{10.5555/3666122.3667065} and Copilot \cite{ma2023aibetterprogrammingpartner}, can generate code directly from natural language descriptions. 
While there are several works and benchmarks in the literature concerned with method-level code generation, to the best of our knowledge there is only one study and benchmark on class-level code generation \cite{du2023classevalmanuallycraftedbenchmarkevaluating}, and none on diagram-level code generation. Moreover, none of the studies listed in the above surveys focus on code generation leveraging multimodal prompts that include sketches \& diagrams. 

Code summarisation \cite{10.1145/3597503.3639183} aims to automatically generate descriptions of a given source code. This technique improves code comprehension, documentation, and collaboration by providing clear summaries. Existing studies in code summarisation focus on analysing code structures and contexts to generate informative natural language summaries. None of the studies listed in the above surveys focus on code summarisation producing a diagram as its output.

Finally, while there exist a few studies that investigated the generation of UML diagram with support from LLMs \cite{conrardy2024imageumlresultsimage,10664407,Cmara2023}, none of these studies go as far as investigating the generation of working code from UML diagrams, as well as the reverse engineering of relevant UML diagrams from existing code. Perhaps the most interesting study among the three listed is \cite{Cmara2023}, which focuses on building UML class diagrams in PlantUML notation, which we also adopt in this work, by using ChatGPT as a modelling assistant prompted with instructions in natural language.

\section{Methodology}\label{sec:methodology}

We develop a portfolio of novel GenSE use cases that, to the best of our knowledge, have not been previously investigated in the literature.

We utilise Microsoft Copilot in its web-based version, which is based on GPT-4o and allows image attachments. To ensure reproducibility of the discussion in Sections \ref{sec:static_modelling}-\ref{sec:design_patterns}, we have also developed an equivalent pipeline in Gemini, formalised in a Jupyter Notebook based on \texttt{gemini-1.5-flash}, which is included in the supplementary material (SM).\footnote{\url{https://github.com/gwr3n/gense}} In both cases, we left all LLM parameters to their default settings and we did not specify any role or context instructions.

Table \ref{tab:use_cases}  maps use cases discussed in the rest of this work to relevant SE sub-tasks of interest.
All use cases are based on a duly documented interaction with the LLMs that takes the form of a chat comprising multiple rounds of questions and responses (Prompt Chaining), which allow users to step incrementally towards answers and thus get help with multipart problems \cite{gemini_api_docs}. The core principle underpinning all our use cases is to illustrate possible strategies to leverage UML diagrams in order to guide the software development process. More specifically, in Section \ref{sec:static_modelling} we leverage class diagrams to guide the LLM in the context of implementing relevant classes, attributes, and operations; 
in Section \ref{sec:dynamic_modelling} we leverage interaction diagrams to guide the LLM in the context of implementing the desired system behaviour; in Section \ref{sec:hand_drawn} we leverage hand-drawn activity diagrams to guide the LLM in the context of implementing a given method; in Section \ref{sec:design_patterns} we leverage design patterns to influence the design of a given system. Finally, in Section \ref{sec:complex_systems} we present three additional case studies: in the first, we leverage hand-drawn diagrams and design patterns to implement a mathematical expressions evaluator; in the second and third, which feature a higher degree of complexity, we leverage class diagrams to implement a tic tac toe game and a game of checkers, respectively. 

\begin{table}
    \centering
    \begin{tabular}{lrr}
        \toprule
        Use case 				& Section 					& SE sub-tasks\\
        \midrule
	static modelling			& \ref{sec:static_modelling}	&SD/DE/CS\\
        dynamic modelling		& \ref{sec:dynamic_modelling}	&SD/DE/CS\\
        hand drawn diagrams	& \ref{sec:hand_drawn}		&DE/CS\\
        design patterns			& \ref{sec:design_patterns}	&SD\\
        expression evaluator		& \ref{sec:complex_systems}	& SD/DE\\
        tic tac toe				& \ref{sec:complex_systems}	& SD/DE\\
        checkers 				& \ref{sec:complex_systems}	& SD/DE\\
        \bottomrule
    \end{tabular}
    \caption{Use cases investigated in the rest of this work (SD: software design; DE: software development; CS: code summarisation)}
    \label{tab:use_cases}
\end{table}

\section{Multimodal GPTs in software development}\label{sec:GenSE}

In what follows, we will assume that a preliminary requirement analysis has been carried out, which has produced a portfolio of initial user stories. While GPTs may in principle support automated elicitation of user stories and fully automated translation of user stories into working software, as things stand today this level of automation in the realm of code generation is hardly found in SE practice; nor we suspect it would be beneficial, as it may conflict with some of the principles of Agile. What we find in practice are teams of software developers collaborating to translate user stories into working software. Our focus is to illustrate novel use cases of GPTs in this specific context. 

The role of sketches and diagrams in the daily work of software developers has been investigated in \cite{Baltes2014}. This study found that most practitioners produced informal hand-drawn diagrams and did not use UML; those using UML, tended to use it informally and selectively. 
We argue that this stems from the fact that the adoption of a ``formal'' notation, at present, bears no advantage with respect to an ``informal'' one.
Over several decades, firms have repeatedly tried to develop tools (e.g. IBM Rational) that could automatically generate working code from UML diagrams, or reverse engineer UML diagram from existing code. While these technologies still exist, it is rare to find developers who routinely develop a complete set of UML diagrams and then translate this to code using such tools. The lacklustre success of these tools is likely due to a fundamental misunderstanding of what UML is: a language for capturing and communicating conceptual requirements, not one aimed at describing a complete system. In other words, UML diagrams are most useful when they are high level and sufficiently abstract. Generation of complete working code requires such diagrams to reach a level of detail that is equivalent to the working code itself; but generating diagrams at this level of abstraction would be a complete waste of time: why then not generating the code itself directly?

Albeit there are already good reasons for practitioners to produce sketches and diagrams in certain circumstances, we believe the advent of multimodal GPTs, such as GPT-4 \cite{openai2024gpt4technicalreport}, will substantially increase the associated use cases. To exemplify this, in what follows we outline a set of use cases illustrating how practitioners may combine multimodal GPTs and UML diagrams to innovate SE practices. 

\subsection{Static modelling}\label{sec:static_modelling}

\begin{figure}[h]
    \centering
    \includegraphics[width=0.8\columnwidth]{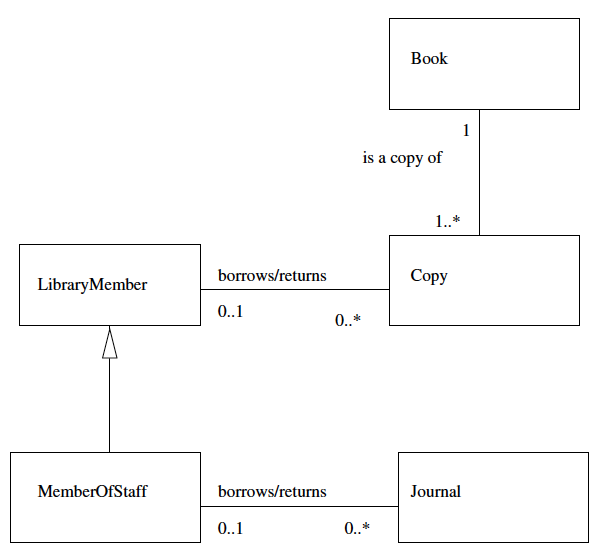}
    \caption{Class model of a library}
    \label{fig:class_diagram}
\end{figure}

As a motivating example to illustrate our use cases, we consider the introductory case study in Chapter 3 of \cite{DBLP:books/daglib/0016253}. In this section, we focus on static modelling; in particular, we assume that the developer has already converted the relevant requirements into a preliminary class diagram such as that shown in Figure \ref{fig:class_diagram}. 
The developer now wants to obtain a preliminary code that matches this class diagram. Rather than using a UML diagram to code conversion tool, we will here rely on Copilot {\em prompting with images}. To achieve this, we attach Figure \ref{fig:class_diagram} and use the prompt illustrated in the following greyed box.
\begin{tcolorbox}[enhanced jigsaw,colback=black!10!white, boxrule=0.2mm]
Implement the attached UML class diagram in Python.
\end{tcolorbox}
The resulting code is presented in the SM. It is noteworthy that in addition to implementing the classes, Copilot also made an attempt at drafting part of the business logic by leveraging semantic information associated with the labels of the relationships in the diagram. 

\begin{figure}[h]
    \centering
    \includegraphics[width=\columnwidth]{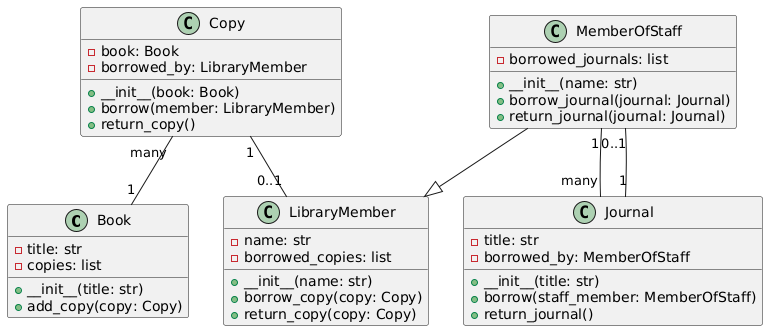}
    \caption{The reconstructed class diagram}
    \label{fig:reconstructed_class_diagram}
\end{figure}

To validate the classes generated against the original UML diagram, we can use Prompt Chaining and ask Copilot to generate a corresponding class diagram in PlantUML\footnote{https://www.plantuml.com/} notation via the following prompt.
\begin{tcolorbox}[enhanced jigsaw,colback=black!10!white, boxrule=0.2mm]
Develop a class diagram in PlantUML notation for the classes in the Python code generated.
\end{tcolorbox}
%
%
%
%
%
%
We can then use PlantUML to visualise the corresponding UML class diagram, which is shown in Figure \ref{fig:reconstructed_class_diagram} and, incidentally, only partly matches the original design in Figure \ref{fig:class_diagram}. This is due to the fact that the Python implementation, whilst not incorrect, does not fully capture all cardinality constraints in the original diagram.

\subsection{Dynamic modelling}\label{sec:dynamic_modelling}

\begin{figure}[h]
    \centering
    \includegraphics[width=\columnwidth]{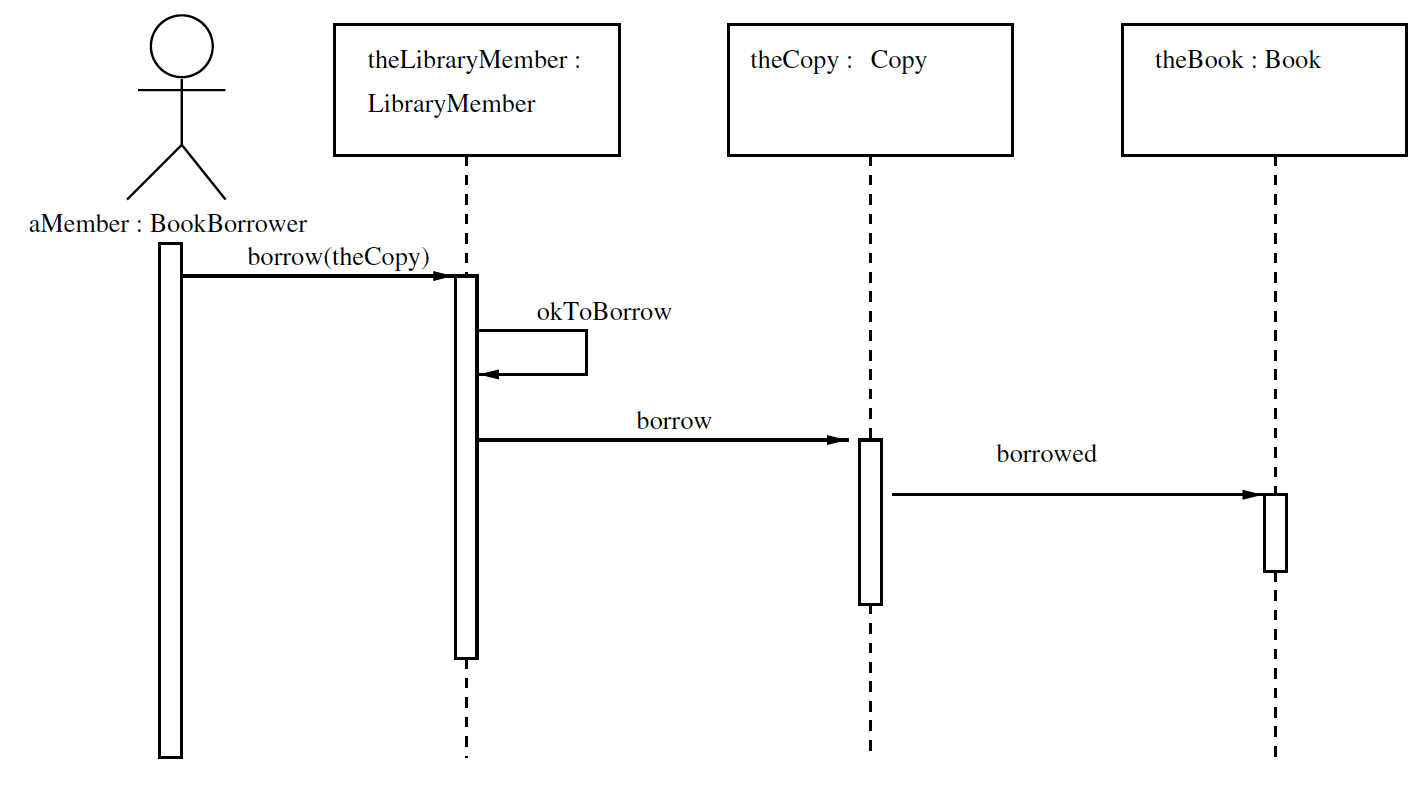}
    \caption{A sequence diagram illustrating an interaction}
    \label{fig:sequence_diagram}
\end{figure}

We may now want to proceed and describe the dynamic behaviour of the system. Rather than using natural language, we may describe specific aspect of such behaviour by leveraging suitable UML interaction diagrams, such as the {\em sequence diagram} in Figure \ref{fig:sequence_diagram}.
To achieve this, we will attach Figure \ref{fig:sequence_diagram} and chain the following prompt in Copilot.
\begin{tcolorbox}[enhanced jigsaw,colback=black!10!white, boxrule=0.2mm]
Implement the dynamic behaviour illustrated in the attached sequence diagram. Do not explicitly implement the actor ``BookBorrower.''
\end{tcolorbox}
\noindent
The resulting code is presented in the SM.

\begin{figure}[h]
    \centering
    \includegraphics[width=0.8\columnwidth]{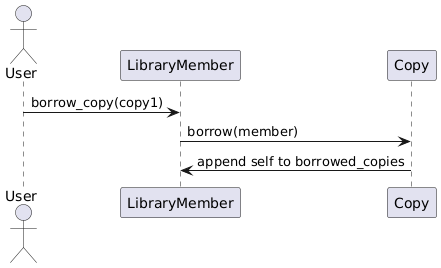}
    \caption{A communication  diagram}
    \label{fig:communication_diagram}
\end{figure}

Next, we may want to explore the behaviour of the various entities involved in a method call by obtaining a {\em communication diagram} for it. Copilot can generate this diagram (Figure \ref{fig:communication_diagram}) via the following prompt chained to previous outputs.
\begin{tcolorbox}[enhanced jigsaw,colback=black!10!white, boxrule=0.2mm]
Generate a UML communication diagram in PlantUML notation illustrating the behaviour of the following code: member.borrow\_copy(copy1).
\end{tcolorbox}
%
%

\begin{figure}[h]
    \centering
    \includegraphics[width=0.4\columnwidth]{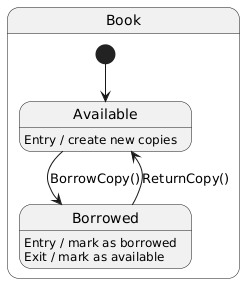}
    \caption{The reconstructed state diagram}
    \label{fig:state_diagram}
\end{figure}

In the code generated by Copilot, the state of the book object changes when a copy of the book is successfully borrowed. In particular, the book may change from being borrowable (there is a copy of it in the library) to not borrowable (all copies are out on loan or reserved). This behaviour can be represented via a {\em state diagram}. Rather than drawing such state diagram, we generate one via the following prompt.
\begin{tcolorbox}[enhanced jigsaw,colback=black!10!white, boxrule=0.2mm]
Generate a UML state diagram in PlantUML notation to represent the possible states of the Book object.
\end{tcolorbox}
%
%
%
%
\noindent
This leads to the state diagram in Figure \ref{fig:state_diagram}.

\begin{figure}[h]
    \centering
    \includegraphics[width=1\columnwidth]{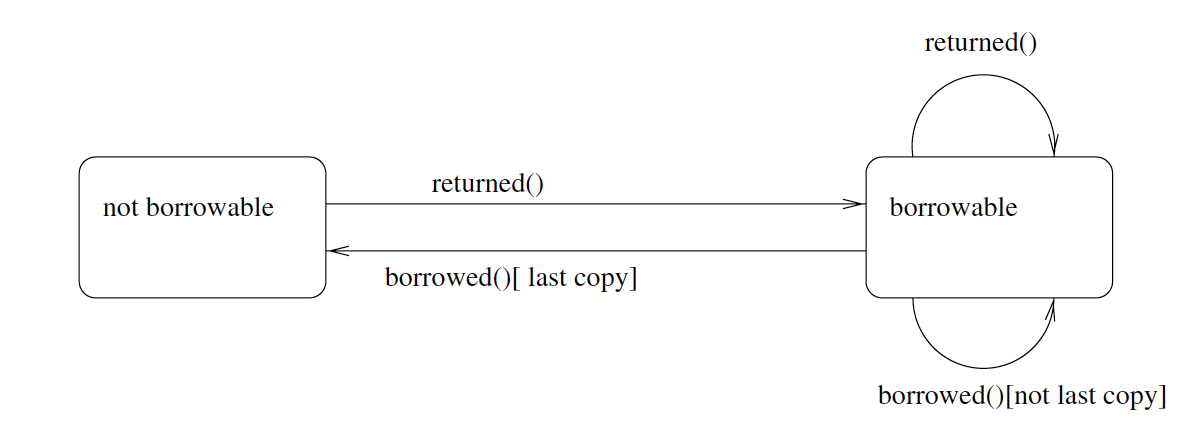}
    \caption{The desired state diagram}
    \label{fig:desired_state_diagram}
\end{figure}

Assume now that the desired behaviour, illustrated in Figure \ref{fig:desired_state_diagram}, is slightly different from that obtained in this preliminary draft of our code.
We can ask Copilot to amend the code by chaining the following prompt.
\begin{tcolorbox}[enhanced jigsaw,colback=black!10!white, boxrule=0.2mm]
Amend the Python classes to capture the behaviour in the attached state diagram. Make sure the generated code continues to reflect the original class diagram.
\end{tcolorbox}
\noindent
The resulting code is presented in the SM.

\begin{figure}[h]
    \centering
    \includegraphics[width=0.7\columnwidth]{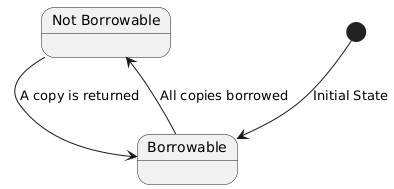}
    \caption{The revised state diagram}
    \label{fig:revised_state_diagram}
\end{figure}

Finally, we can ask Copilot to generate a new state diagram in PlantUML notation that reflects the behaviour of the updated code. The resulting diagram is shown in Figure \ref{fig:revised_state_diagram} and matches the behaviour in Figure \ref{fig:desired_state_diagram}.


\subsection{Hand-drawn diagrams}\label{sec:hand_drawn}

\begin{figure}[h]
    \centering
    \includegraphics[width=1\columnwidth]{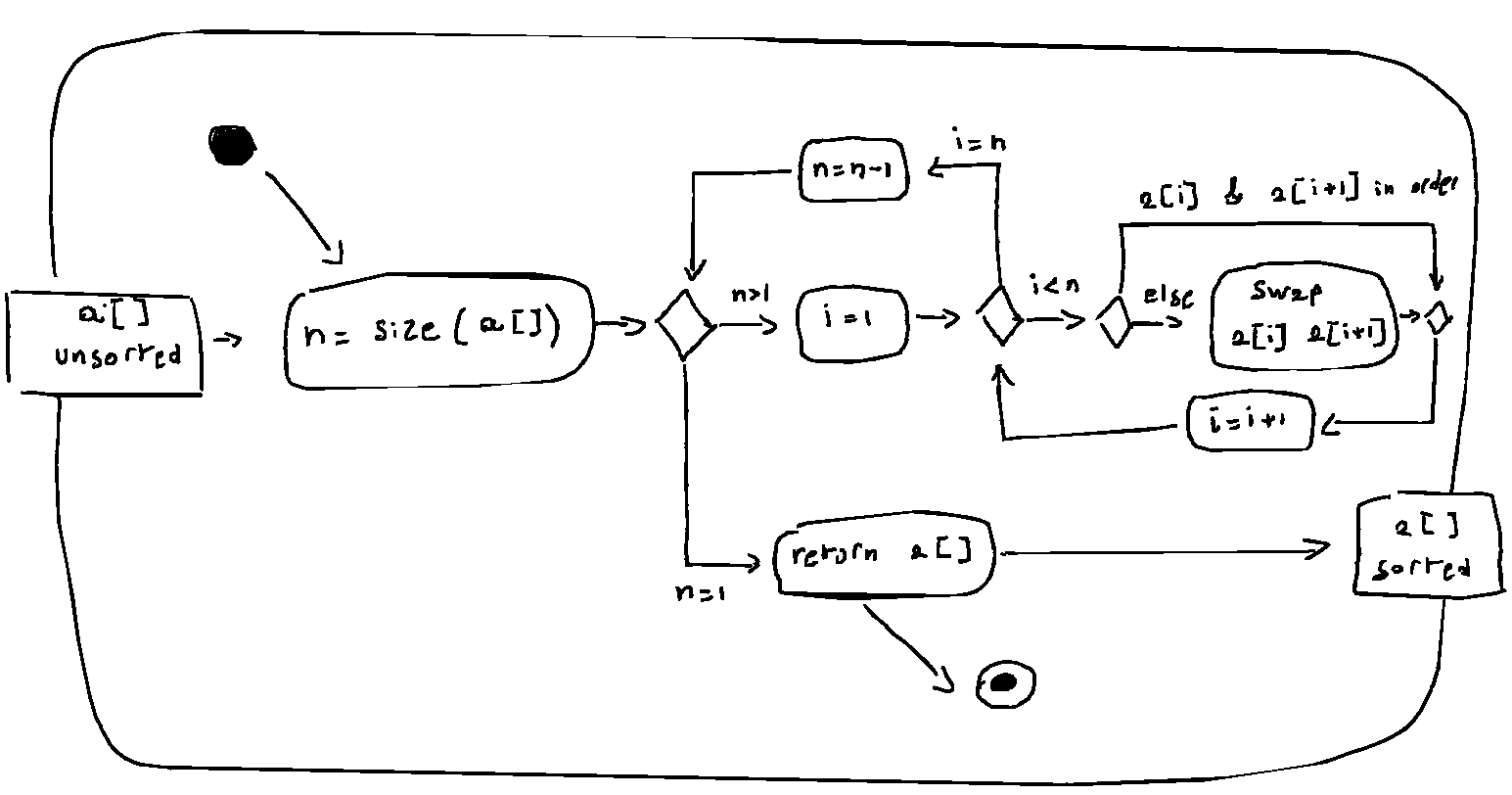}
    \caption{An activity diagram}
    \label{fig:activity_diagram}
\end{figure}

While all the previous examples featured computer generated diagrams, Copilot is able to handle equally well hand-drawn diagrams. We consider the {\em activity diagram} shown in Figure \ref{fig:activity_diagram}. 
We attach the diagram, and input the following prompt.
\begin{tcolorbox}[enhanced jigsaw,colback=black!10!white, boxrule=0.2mm]
Create a method in Python that implements the attached UML activity diagram.
\end{tcolorbox}
Copilot correctly recognises the fact that that the activity diagram represents a sorting algorithm (bubble sort), and returns a method implementing it.

\begin{figure}[h]
    \centering
    \includegraphics[width=0.58\columnwidth]{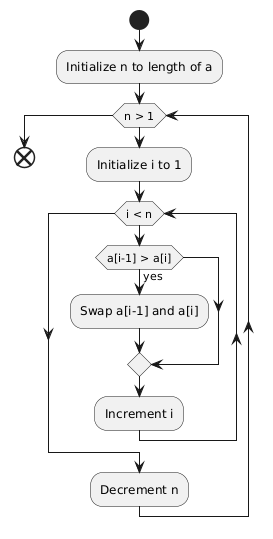}
    \caption{A PlantUML activity diagram of a sorting algorithm}
    \label{fig:activity_diagram_plantuml}
\end{figure}

Alternatively, assuming the code for our sorting algorithm is already available to us, we can ask Copilot to convert it to an activity diagram via the following prompt.
\begin{tcolorbox}[enhanced jigsaw,colback=black!10!white, boxrule=0.2mm]
Create an activity diagram in PlantUML notation for the following method in Python: [insert Python code].
\end{tcolorbox}
In this specific instance, Copilot returned a diagram with a small number of syntax errors. The errors were minor issues with the PlantUML syntax of the two \texttt{while} loops in the diagram, and could be easily fixed by hand. The resulting activity diagram is shown in Figure \ref{fig:activity_diagram_plantuml}. Conversely, a second prompt asking Copilot to generate a diagram in PlantUML notation by translating directly the diagram in Figure \ref{fig:activity_diagram} produced no errors. Since the focus of our discussion is {\em facilitating} --- and not fully automating --- SE activities, the presence of minor errors is of no concern, as these can easily be addressed in follow up prompts.

\subsection{Design patterns} \label{sec:design_patterns}

\begin{figure}[h]
    \centering
    \includegraphics[width=0.7\columnwidth]{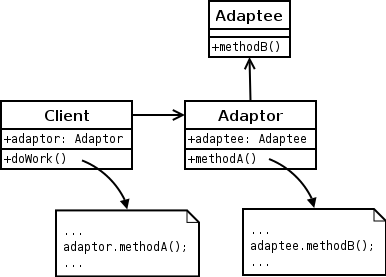}
    \caption{The ``Adapter'' Pattern}
    \label{fig:adapter_pattern}
\end{figure}

Design Patterns \cite{Gamma1994-br} are reusable solutions to commonly occurring problems in software design. They are not specific implementations but rather general templates that can be adapted to various situations. By using design patterns, developers can create more flexible, maintainable, and efficient software systems. There exists a wealth of existing patterns available in the SE literature; In addition, new patterns can be created and illustrated via appropriate UML diagrams. For instance, we may consider the ``Adapter'' pattern, which converts the interface of a class into another interface clients expect. This pattern is illustrated in Figure \ref{fig:adapter_pattern}. 
Design patterns offer endless opportunities to carry out Few-Shot and CoT prompting. We shall illustrate this point with a practical example involving the ``Adapter'' pattern.

%
%
%
 \begin{figure}[h]
    \centering
    \includegraphics[width=0.8\columnwidth]{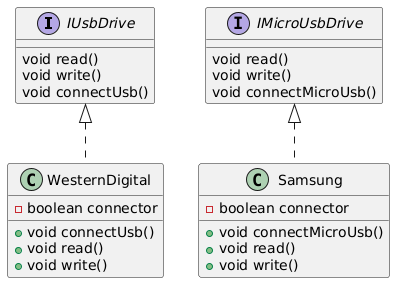}
    \caption{The data transfer system}
    \label{fig:data_transfer}
\end{figure}
\begin{figure}[h]
    \centering
    \includegraphics[width=0.9\columnwidth]{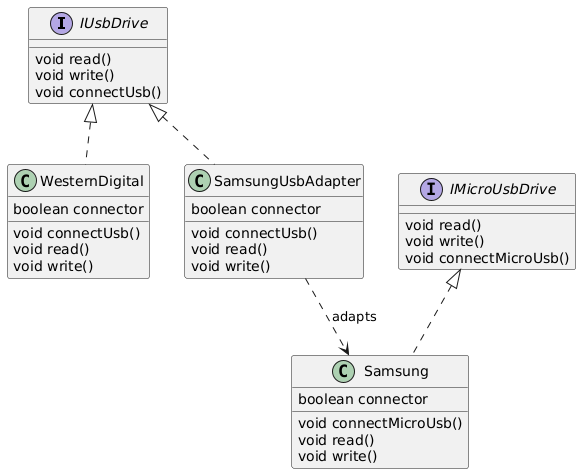}
    \caption{The data transfer system with an adaptor.}
    \label{fig:data_transfer_adapted}
\end{figure}

Consider the class diagram in Figure \ref{fig:data_transfer}. In this system we have two portable drives, a Samsung drive and a Western Digital drive. The Samsung drive features a microusb connector, while the Western Digital drive features a traditional usb connector. Our aim is to allow the Samsung drive to read and write data via usb. To achieve this, we attach the diagram to Copilot and provide the following prompt.
\begin{tcolorbox}[enhanced jigsaw,colback=black!10!white, boxrule=0.2mm]
Extend the UML diagram by using the Adapter design patter to allow a Samsung drive to read and write data via Usb. Provide the output in PlantUML notation.
\end{tcolorbox}
The resulting diagram, which correctly implements the Adapter pattern, is illustrated in Figure \ref{fig:data_transfer_adapted}.
%
%
%
%
%
%
Since the Adapter pattern is well known, and there is no need to provide a diagram for it while prompting. 
However, if the user intends to utilise a brand new pattern, then Few-Shot prompting can be leveraged to obtain the desired result.
In practice, this would entail attaching a UML diagram representing the new design pattern that needs to be operationalised in the context of the given coding task.

\subsection{Implementation of complex systems} \label{sec:complex_systems}

In this section, we discuss the implementation of three more complex applications. In all cases, the interaction takes the form of a chat comprising multiple rounds of questions and responses (Prompt Chaining), which allows the user to step incrementally towards a working implementation.

\subsubsection{Evaluating mathematical expressions}

In this first application, we consider a class diagram (provided in our SM) that conceptually captures a system that evaluates mathematical expressions comprising additions and subtractions. The system comprises five interrelated classes. We were able to obtain a complete implementation with a single multimodal prompt (diagram + instruction to implement the diagram). An additional prompt was required to implement an evaluator by leveraging the Visitor design pattern.

\subsubsection{Tic tac toe}

We consider the class diagram\footnote{\url{https://github.com/pelensky/JavaTTT/blob/master/UML.pdf}} for a tic tac toe game. The diagram comprises over twenty interrelated classes. We use this diagram to guide the development of a tic tac toe Python application. The detailed interaction with Copilot is illustrated in our SM. This interaction involves seven prompts in which Copilot is asked to fully implement specific classes in the diagram. In line with traditional SE practices, we start from classes featuring low complexity and connectivity, and we then move towards other classes that depend on those already implemented. This is followed by two prompts to correct errors encountered while trying to execute the application. In each of these prompts the full stack trace of the error is fed to Copilot. Finally, an additional prompt is used to tailor board visualisation. A complete application that neatly matches the conceptual model presented in the class diagram is hence obtained in ten short prompts comprising a single sentence each.

\subsubsection{Checkers}

We consider the class diagram\footnote{\url{https://app.genmymodel.com/api/repository/wanex505/checkers}} for a checkers game. The diagram comprises over ten interrelated classes and a complex underpinning logic. We use this diagram to guide the development of a checkers Python application. The detailed interaction with Copilot is illustrated in our SM. This interaction involves nine prompts in which Copilot is asked to fully implement specific classes in the diagram. This is followed by an additional prompt to develop a text-based drawing logic. Ten additional prompts are then required to correct errors encountered while trying to execute the application and play the game. These errors include both exceptions (e.g. a class attribute is missing) as well as conceptual errors that affect the game logic (a piece is not removed from the board after being jumped over) or the user interaction (e.g. the user is requested to input a move, but it is not stated if they are playing as black or white). A complete application that matches the conceptual model presented in the class diagram is hence obtained in twenty short prompts comprising a single sentence each.

In our SM we provide a more detailed description of each case study, as well as Jupyter Notebooks reporting the complete interactions with Copilot.

\section{Discussion and conclusions}\label{sec:discussion}

In this section, we reflect on how the use cases discussed in Section \ref{sec:GenSE}, which are made possible thanks to the enhanced capabilities of GPT-4, address several research gaps in GenSE, as well as some long standing issues in SE. We will also discuss how the same use cases open new research directions that can be investigated in future studies. 

We shall next focus on the research gaps we bridged in GenSE.
First, our study addresses the lack of applications of GenAI to the software design sub-task of SE, a literature gap identified in \cite{huang2024generativesoftwareengineering,10.1145/3695988}. 
Second, to the best of our knowledge, our study represents the first application of multimodal GPTs (diagram + prompt) to the software development sub-task of SE, which in GenSE is generally carried out via natural language-based prompts. We argue that diagram-based prompting may be advantageous, in terms of information transfer efficiency, compared to pure natural language prompting, as diagrams leverage background knowledge associated with their visual elements and structure, thereby leading to compression \cite{li2024understandingcompression}. 
Third, by reverse engineering code into diagrams, we have contributed to the code summarisation sub-task of SE, which is again predominantly carried out via natural language summaries. 
Fourth, while there exist studies that investigated the production of PlantUML code from hand-drawn diagrams \cite{conrardy2024imageumlresultsimage,10664407}, we believe this is the first study in which we cover multiple elements of the SE stack, from conceptual modelling --- which includes static and dynamic analysis of the software system --- to implementation. Moreover, to the best of our knowledge, this is the first study that focus on diagram-level (rather than method-level or class-level) code generation.
Finally, the use of existing (or new) design patterns in the context of Few-Shot prompting for software system design does not seem to have been investigated before in the literature, and hence merits attention in future studies.

Next, we shall look at which long standing issues in SE may be addressed by the use cases discussed in our work.
In \cite{Baltes2014}, the authors suggested that sketches could supplement often outdated and poorly written documentation, advocating for tools to archive and retrieve these sketches. No such tool exists and poorly documented code remains a persistent issue in SE. However, GenAI has already started to change this picture via automatic generation of code comments. And yet, if sketches and diagram (both formal and informal) can be transformed, as we have shown, into a preliminary draft of the desired source code, then producing reliable visual descriptions of the software system under development suddenly becomes appealing. Likewise, if it is possible to selectively and cheaply reverse engineer part of such software system to visually inspect the behaviour of classes and methods, the developer becomes equipped with a formidable arsenal of {\em high resolution} communication tools to enhance code understanding, boost team communication, and resolve the long lasting conflict between investing time in developing working software or writing comprehensive documentation. These are all important and novel directions of enquiry that should be investigated in future studies.

\newpage

\bibliographystyle{named}
\bibliography{gense}

\end{document}